\documentclass [12pt,article,aps,showpacs,showkeys,floatfix]{revtex4}
\def\be{\begin{eqnarray} &&}

\def\ee{\end{eqnarray}}

\usepackage{graphicx}
\usepackage{epsfig}
\usepackage{amssymb}
\usepackage{amsmath}

\begin{document}

\newcommand*{\Pavia}{Dipartimento di Fisica Nucleare e Teorica,
Universit\`a degli Studi di Pavia, Pavia, Italy}\affiliation{\Pavia}
\newcommand*{\INFN}{Istituto Nazionale di Fisica Nucleare,
Sezione di Pavia, Pavia, Italy}\affiliation{\INFN}
\newcommand*{\Mainz}{Institut f\"ur Kernphysik, Johannes Gutenberg Universit\"at,
D-55099 Mainz, Germany}\affiliation{\Mainz}
\title{Higher order forward spin polarizability}

\author{B.~Pasquini$^{a,b}$, P.~Pedroni$^b$, D.~Drechsel$^c$}
\affiliation{$^a$Dipartimento di Fisica Nucleare e Teorica, Universit\`a degli Studi di Pavia, Pavia, Italy\\
$^b$Istituto Nazionale di Fisica Nucleare, Sezione di Pavia, Pavia, Italy\\
$^c$Institut f\"ur Kernphysik, Johannes Gutenberg Universit\"at, D-55099 Mainz, Germany\\}


\begin{abstract}
As a guideline for future experiments to extract the four (leading) spin polarizabilities of the nucleon, we have constructed
the forward amplitude for polarized Compton scattering by dispersion integrals.
These integrals have been saturated by recently measured helicity-dependent photoabsorption cross sections
as well as predictions for pion photoproduction multipoles from several phenomenological
descriptions and chiral perturbation theory. The comparison of these results corroborates the strategy
to extract the spin polarizabilities by fitting them to polarized Compton data
and fixing all higher order spin effects by dispersion relations based on pion photoproduction multipoles.
\end{abstract}

\maketitle
\begin{titlepage}
\maketitle
\end{titlepage}
\section{Introduction}
Real Compton scattering (RCS) probes the response of a complex system to an external electromagnetic field.
In particular, photon scattering off the nucleon is described by
six independent helicity amplitudes depending on the lab energy $E_{\gamma}$ and the scattering
angle $\theta$ of the photon. The low-energy expansion (LEX) of these amplitudes defines
two spin-independent (scalar) and four spin-dependent (vector) polarizabilities, appearing at order
$E_{\gamma}^2$ and $E_{\gamma}^3$, respectively.
These polarizabilities are fundamental structure constants of the nucleon just
as its size and shape.\\
Whereas the scalar polarizabilities have been measured quite accurately by comparing the RCS
data directly to the LEX~\cite{OlmosdeLeon:2001zn,Schumacher:2005an}, the quest to determine the vector polarizabilities is still going on.
The difficulty to measure the vector polarizabilities is caused by their suppression at low energies,
which makes it impossible to extract them at energies $E_{\gamma} < 100$~MeV. With increasing
energies, on the other hand, the LEX converges slower and slower and finally breaks down at
pion-production threshold, $E_{\gamma}=m_{\pi}+m_{\pi}^2/(2M)\approx 150$~MeV,
where $m_{\pi}$ and $M$ are the pion and nucleon masses, respectively. In order to determine the
vector polarizabilities from RCS data, it is prerequisite to
(i) analyze the data within a theoretical framework covering the energy region up to the
$\Delta$(1232) resonance at $E_{\gamma} \approx 300$~MeV and
(ii) perform dedicated experiments with polarized photons and nucleons with increased
sensitivity to the spin-dependent amplitudes.\\
The sensitivity of various single and double polarization observables to the
spin polarizabilities has been studied by subtracted dispersion relations
based on the pion photoproduction multipoles~\cite{Pasquini:2007hf}. The authors concluded
that a complete separation of the polarizabilities should be possible by dedicated
polarization experiments between pion threshold and the $\Delta$(1232) region,
provided that polarization measurements can be achieved within an accuracy of a few percent.
Such an  experiment to extract the spin polarizabilities has recently been proposed
at MAMI~\cite{MAMI:2009dd}.\\
The predictions of chiral perturbation theory (ChPT) at 
${\mathcal {O}}(p^4)$ are in good agreement with the
empirical data for the scalar polarizabilities of the 
proton~\cite{Bernard:1993ry}. However, in order to cover the energy region
necessary for the extraction of the vector polarizabilities, the 
$\Delta$(1232) must be included
as a dynamic degree of 
freedom~\cite{Hemmert:1997tj,Hildebrandt:2003dd,Hildebrandt:2003md,Pascalutsa:2002pi}.
\newline
\noindent
Recently, these polarizabilities have also been studied in lattice QCD. 
The first results for the magnetic polarizability of the proton
are quite encouraging~\cite{Lee:2005dq}, whereas quite different values 
are reported for the electric polarizability
~\cite{Christensen:2004ca,Detmold:2010ts,Shintani:2006xr,Engelhardt:2007ub,Alexandru:2008sj,Guerrero:2009yv}.
It was also proposed to extract the spin polarizabilities from lattice 
calculations~\cite{Detmold:2006vu}.
However, calculations with dynamic and lighter quarks are prerequisite 
to extrapolate to the physical point in a reliable way,
because these polarizabilities show distinct signatures of chiral 
dynamics~\cite{Hildebrandt:2003dd}.
\\
As stated above, the complete separation of the six polarizabilities of the nucleon requires both
new polarization data and a theoretical framework for RCS, which will necessarily contain some
model dependence through parameters describing the high-energy regime. Such parameters are the low-energy
constants appearing in ChPT, and in dispersion relations the extrapolation of the photoproduction
data to regions not covered by the experiment. Any additional confirmation of the theoretical framework by
independent experimental information is therefore most welcome. On that ground it is the aim of the present
work to construct the full spin-dependent RCS amplitude in forward direction from data
given by (helicity-dependent) total photoabsorption cross sections. In Sec. II we present the formalism
necessary to discuss the forward scattering amplitudes and their relations to dispersion integrals over
inclusive photoabsorption cross sections, the LEX defining the leading and higher order
forward spin polarizabilities (FSPs) as well as the
``dynamic'' FSP describing the full spin-dependent response to forward scattering. The relevant dispersion integrals
are evaluated in Sec. III and compared with several model predictions. Finally, we conclude with a brief section
summarizing our findings.
\section{Kinematics, amplitudes, and spin polarizabilities}
The polarizability of the nucleon is determined by Compton scattering~\cite{Babusci:1998ww,Drechsel:2002ar},
\begin{equation}
\gamma(k)+N(p)\rightarrow\gamma(k')+N(p')\,,
\label{eq:RCS}
\end{equation}
where $k$ and $p$ denote the momentum four-vectors of the incoming photon and nucleon,
respectively, with the primed quantities standing for the final state momenta.
Compton scattering can be described by the two Lorentz invariants $\nu=K \cdot P/M$ and $t=(k-k')^2$,
with $K=(k+k')/2$ and $P=(p+p')/2$. These invariants are related to
the initial ($E_{\gamma}$) and final ($E_{\gamma}'$) photon laboratory energies, and to
the laboratory scattering angle $\theta_{\mathrm{lab}}$ by
\begin{align}
t &  =  - 4E_{\gamma}E_{\gamma}'\sin^2 \frac{\theta_{\mathrm{lab}}}{2} =
- 2M (E_{\gamma}-E_{\gamma}')\ , \nonumber \\
\nu & =  E_{\gamma} + \frac{t}{4M} = \frac{1}{2}
(E_{\gamma}+E_{\gamma}').
\label{eq:kin-lab}
\end{align}
In the forward direction, the invariants take the values
\begin{equation}
\nu = E_{\gamma} = E_{\gamma}' \, , \quad t=0.
\label{eq:kin-lab2}
\end{equation}
For the theoretical calculations of Compton scattering it is convenient to use the center-of momentum (c.m.) system.
With $\omega$ and $\omega'$ the photon c.m. energies in the initial and final states, respectively,
and $\omega = \omega'$, the analog of Eq.~(\ref{eq:kin-lab2}) in the c.m. frame is
\begin{equation}
\nu =  \omega (\sqrt {1+\omega^2/M^2} +\omega/M) \, , \quad t=0.
\label{eq:kin-cm}
\end{equation}
The forward Compton amplitude takes the form
\begin{equation}
T(\nu,\,\theta=0) = {\vec{\varepsilon}}\,'^{\ast}\cdot\vec{\varepsilon}\,f(\nu)+
i\,\vec{\sigma}\cdot({\vec{\varepsilon}}\,'^{\ast}\times\vec{\varepsilon})\,g(\nu)\, ,
\label{eq:amplitude}
\end{equation}
where $\vec{\varepsilon}$ and $\vec{\varepsilon}\,'$ are the photon polarizations
in the initial and final states, respectively, and $\vec{\sigma}$ is the spin operator
of the nucleon. Because of the crossing symmetry, the amplitude $T$ is invariant
under the transformation $\vec{\varepsilon} \rightarrow \vec{\varepsilon}\,'$ and
$\nu \rightarrow -\nu$. As a result, $f$ is an even and $g$ an odd function of $\nu$.\\
The amplitudes $f$ and $g$ can be determined by scattering circularly
polarized photons (e.g., helicity $\lambda=1$) off nucleons
polarized along or opposite to the photon momentum $\vec {k}$. Depending on the relative
orientation of the spins, the absorption of the photon leads to hadronic excited states
with spin projections $1/2$ or $3/2$. The optical theorem expresses the unitarity of the
scattering matrix by relating the respective absorption cross sections,  $\sigma_{1/2}$
and $\sigma_{3/2}$, to the imaginary parts of the forward scattering amplitudes,
\begin{align}
{\rm{Im}}\ f(\nu) &= \frac{\nu}{8\pi}
\bigg( \sigma_{1/2}(\nu)+\sigma_{3/2}(\nu) \bigg)\,, \nonumber\\
{\rm{Im}}\ g(\nu) &= \frac{\nu}{8\pi}
\bigg( \sigma_{1/2}(\nu)-\sigma_{3/2}(\nu) \bigg)\,.
\label{eq:opt-theorem}
\end{align}
In the following, we restrict the discussion to the spin-dependent amplitude $g(\nu)$.
Using causality, the crossing symmetry, the optical theorem and an appropriate high-energy
behavior of the scattering amplitude, we may set up the following unsubtracted dispersion relation:
\begin{equation}
{\rm{Re}}\ g(\nu) = \frac{\nu}{4\pi^2}\,{\mathcal{P}}\,
\int_{\nu_0}^{\infty}\,\frac{\sigma_{1/2}(\nu')-\sigma_{3/2}(\nu')}
{\nu'^2-\nu^2}\,\nu'd\nu'\, ,
\label{eq:DR}
\end{equation}
where $\nu_0=m_{\pi}+m_{\pi}^2/(2M)$ is the threshold for producing a pion
with mass $m_{\pi}$. If the integral exists, it has 
a Taylor series expansion about $\nu=0$, with a convergence radius given by the onset
of inelasticities at $\nu_0$. This series can be compared to the
LEX of the spin-flip amplitude~\cite{Drechsel:2002ar},
\begin{equation}
g(\nu)=-\frac{e^2\kappa_N^2}{8\pi M^2}\nu+\gamma_0\nu^3+ \bar\gamma_0\nu^5+{\cal O}(\nu^7).
\label{eq:LEX}
\end{equation}
The leading term of this expansion is due to intermediate nucleon states (Born terms).
The comparison of the LEX with the Taylor expansion of Eq.~(\ref{eq:DR}) yields the
Gerasimov-Drell-Hearn (GDH) relation between the anomalous magnetic moment of the nucleon,
$\kappa_N$, and the spin-dependent absorption spectrum,
\begin{equation}
\frac{\pi e^2\kappa_N^2}{2 M^2}=\,
\int_{\nu_0}^{\infty}\,\frac{\sigma_{3/2}(\nu')-\sigma_{1/2}(\nu')}
{\nu'}\,d\nu'\, \equiv I_{\mathrm {GDH}}.
\label{eq:GDH}
\end{equation}
The higher-order terms are produced by the hadronic excitation spectrum (non-Born or
dispersive contributions). These terms parameterize the FSPs
of the nucleon. In particular, the leading and next-to-leading FSPs are given by
\begin{eqnarray}
\gamma_0&=&\frac{1}{4\pi^2}\,
\int_{\nu_0}^{\infty}\,\frac{\sigma_{1/2}(\nu')-\sigma_{3/2}(\nu')}
{{\nu'}^3}\,d\nu'\,, \nonumber\\
\bar{\gamma}_0&=&\frac{1}{4\pi^2}\,
\int_{\nu_0}^{\infty}\,\frac{\sigma_{1/2}(\nu')-\sigma_{3/2}(\nu')}
{{\nu'}^5}\,d\nu'\,.
\label{eq:FSP}
\end{eqnarray}
Furthermore, we may define the crossing-even ``dynamic'' FSP $\gamma_0^{\rm{dyn}}(\nu)$ by
\begin{align}
g(\nu)&=-\frac{e^2\kappa_N^2}{8\pi M^2}\nu+\gamma_0^{\rm{dyn}}(\nu)\nu^3 \, ,\nonumber\\
{\rm {Re}}[\gamma_0^{\rm{dyn}}(\nu)] &= \frac{1}{4\pi^2}\,{\mathcal{P}}\,\int_{\nu_0}^{\infty}\,\frac{\sigma_{1/2}(\nu')-\sigma_{3/2}(\nu')}
{\nu' ({\nu'}^2-\nu^2)}\,d\nu' \, ,\label{eq:dyn-FSP}
\\
{\rm {Im}}[\gamma_0^{\rm{dyn}}(\nu)] &= \frac{\sigma_{1/2}(\nu)-\sigma_{3/2}(\nu)}{8 \pi \nu^2} \, .\nonumber
\end{align}
For $\nu<\nu_0$ the imaginary part vanishes, and the ``dynamic'' FSP has the following LEX:
\begin{equation}
\gamma_0^{\rm{dyn}}(\nu)=\gamma_0 + \bar{\gamma}_0\, \nu^2 + {\cal O}(\nu^4).
\label{eq:dyn-FSP-LEX}
\end{equation}
In comparing with the literature, we note that the expression ``dynamic polarizability'' in
Refs.~\cite{Hildebrandt:2003dd} and \cite{Drechsel:2002ar} refers to the energy dependence of the polarizability
due to individual electromagnetic multipole radiation or combinations thereof. On the other hand, the FSP $\gamma_0^{\rm{dyn}}(\nu)$
of Eq.~(\ref{eq:dyn-FSP}) describes the nucleon's forward spin response to the full
electromagnetic field, including the contributions of all the multipoles and their retardation.

The general theory of Compton scattering involves 6 crossing-even amplitudes $A_i(\nu,t)$ describing the
dispersive part of the scattering matrix~\cite{Babusci:1998ww,Drechsel:2002ar}. In particular, the forward scattering amplitude
$g$ is related to $A_4$ as follows:
\begin{equation}
A_4(\nu,t=0)=\frac {2 \pi M}{\nu^3}\bigg( g(\nu)-\nu g'(0) \bigg) \,.
\label{eq:A4-g}
\end{equation}
Using Eqs.~(\ref{eq:dyn-FSP}) - (\ref{eq:A4-g}) we may cast the polarizabilities into the form
\begin{equation}
\gamma_0 = \frac {a_4}{2 \pi M}, \quad \bar\gamma_0\ = \frac {a_{4, \nu}}{2 \pi M},
\label{eq:gamma-a4}
\end{equation}
with
\begin{equation}
a_4 = A_4(0,0)\, , \quad  a_{4,\nu} = \frac {\partial}{\partial \nu^2} A_4(\nu,0)|_{\nu=0} \, .
\label{eq:a4-a4nu}
\end{equation}
If the invariant $\nu$ is expressed by the c.m. energy $\omega$ according to
Eq.~(\ref{eq:kin-cm}), the LEX of Eq.~(\ref{eq:LEX}) takes the form
\begin{align}
g(\omega)&=-\frac{e^2\kappa_N^2}{8\pi M^2}\omega (1+ \frac {\omega}{M}+
\frac {\omega^2}{2M^2} - \frac {\omega^4}{8M^4})\nonumber\\
&+\gamma_0 \omega^3 (1+\frac {3\omega}{M}+\frac {9\omega^2}{2M^2})+
\bar\gamma_0\omega^5+{\cal O}(\omega^6).
\label{eq:spin-flip-exp-cm}
\end{align}
We observe that the nucleon Born term, proportional to $\kappa_N^2$, contributes a series
of recoil terms in  $\omega/M$. The remaining dispersive terms
contain the spin polarizabilities related to the excited states of the nucleon.
As for the $\nu$ expansion, the
spin polarizability $\gamma_0$ appears as coefficient of the third order term,
now in $\omega^3$. In addition, however, $\gamma_0$ appears also with recoil terms of order
$\omega^4$ and $\omega^5$. Combining all the $\omega^5$ dispersive terms to $\tilde\gamma_0 \, \omega^5$,
we find the relation
\begin{equation}
\tilde\gamma_0 = \bar\gamma_0 +\frac {9}{2M^2}\gamma_0.
\label{eq:5th-order}
\end{equation}
According to Ref.~\cite{Holstein:1999uu}, the ${\cal O}(\omega^3)$
and ${\cal O}(\omega^5)$ FSPs have the following multipole decomposition:
\begin{align}
\gamma_0&=-(\gamma_{E1E1}+\gamma_{M1M1}+\gamma_{M1E2}+\gamma_{E1M2}),
\label{eq:gamma0}\\
\tilde \gamma_0 &=-(\gamma_{E1E1\nu}+\gamma_{M1M1\nu}+\gamma_{M1E2\nu}
+\gamma_{E1M2\nu}\nonumber\\
&+\gamma_{E2E2}+\gamma_{M2M2}+\frac{8}{5}\gamma_{E2M3}+\frac{8}{5}\gamma_{M2E3}).
\label{eq:gamma0-bar}
\end{align}
In  Eq.~(\ref{eq:gamma0-bar}), the first row contains retardation corrections
related to the leading FSP,
whereas the second row shows higher multipole structures probing the
quadrupole and octupole excitation of the system. We note
that the above multipole notation of Ref.~\cite{Drechsel:2002ar} is
related to the nomenclature of Ref.~\cite{Holstein:1999uu} as follows:
\begin{align}
\gamma_{E1E1}=\gamma_{E1},&\quad\gamma_{M1M1}=\gamma_{M1},\nonumber\\
\gamma_{M1E2}=\gamma_{E2},& \quad\gamma_{E1M2}=\gamma_{M2},\nonumber\\
\gamma_{E2E2}=\gamma_{ET},&\quad\gamma_{M2M2}=\gamma_{MT},\nonumber\\
\gamma_{E2M3}=\gamma_{M3},& \quad\gamma_{M2E3}=\gamma_{E3}.
\label{eq:multipoles}
\end{align}
\section{Results for the GDH sum rule and the forward spin polarizabilities}
\subsection{Evaluation of the dispersion integrals}
The dispersion integrals defined in
Eqs.~(\ref{eq:GDH}) and~(\ref{eq:FSP}) were evaluated using the data obtained
by the GDH collaboration at the MAMI (Mainz) and ELSA (Bonn) tagged photon
facilities. The contributions of kinematical regions not covered by the data
were determined on the basis of
various multipole analyses for pion photoproduction, with systematic errors estimated by
comparison of different model predictions.
In particular, we used the work of Hanstein {\emph {et al.}}~(HDT) based on dispersion
relations~\cite{Hanstein:1997tp}, 
the recent version of the SAID09 multipole analysis~\cite{Arndt:2002xv},
the unitary isobar model MAID07~\cite{Drechsel:2007if}, the dynamical DMT model~\cite{DMT:2001},
and the predictions of heavy baryon chiral perturbation theory (HBChPT)
according to Ref.~\cite{Holstein:1999uu}.\\
The experimental data base includes:
\begin{itemize}
\item the helicity-dependent differential cross section
[$({\mathrm {d}}\sigma/{\mathrm {d}}\Omega)_{3/2} - ({\mathrm {d}}\sigma/{\mathrm {d}}\Omega)_{1/2}$]
data for the $n\pi^+$ channel measured at $E_\gamma = (0.18 \pm 0.005)$~GeV
and $E_\gamma = (0.19 \pm 0.005)$~GeV in the angular range $45^{\circ} \leq \theta^{\ast} \leq 109^{\circ}$,
where $\theta^{\ast}$ is the pion emission angle in the c.m. frame~\cite {Ahrens:2004pf},
\item  the  helicity-dependent data for the total inclusive cross section,
$\Delta\sigma = \sigma_{1/2} - \sigma_{3/2}$,
starting at $E_\gamma = (0.204 \pm 0.009)$~GeV and extending to
$E_\gamma = (2.82 \pm 0.09)$~GeV~\cite{Ahrens:2001qt,Dutz:2003mm,Dutz:2004}.
\end{itemize}

As a first step to evaluate the dispersion integrals, the data at
$E_\gamma = (0.18 \pm 0.005)$~GeV and $E_\gamma = (0.19 \pm 0.005)$~GeV were used to obtain the
$n\pi^+$ contribution to $\Delta\sigma$ at these energies and in the measured angular range.
The extrapolation into the unmeasured region of $\theta^{\ast}$ was performed with
the HDT analysis which reproduces the experimental data for E$_{\gamma} \leq 0.3$~GeV quite well,
see for example Fig.~7 of Ref.~\cite {Ahrens:2004pf}.
The error associated with this extrapolation was estimated by comparison
with the  SAID09, MAID07, and DMT multipole analyses. As a result we found an absolute systematic error
of about $\pm 5\%$ for the calculated total cross section value.
\begin{figure}
\begin{center}
\epsfig{file=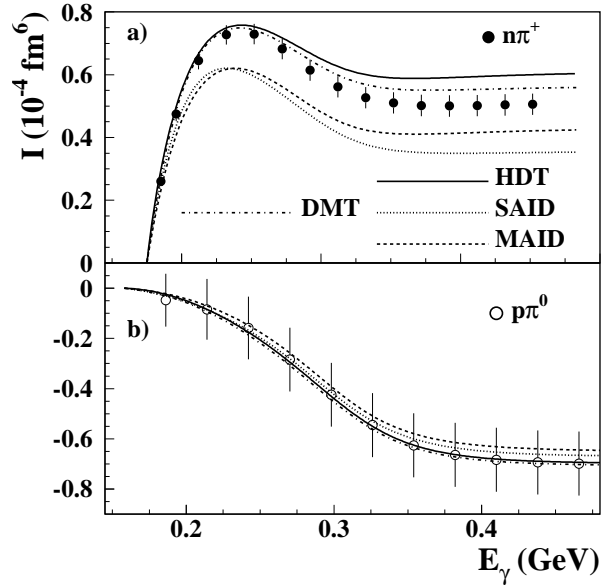,width=0.5\columnwidth}
\caption {The experimental running integrals $I(E_{\gamma})$ of
Eq.~(\protect\ref{exprun}) as obtained from the data of the
GDH Collaboration~\cite{Ahrens:2004pf}
compared to
the HDT~\cite{Hanstein:1997tp}, SAID09~\cite{Arndt:2002xv},  MAID07~\cite{Drechsel:2007if}, and  DMT~\cite{DMT:2001} predictions. The top and bottom panels display
the contributions of the $n\pi^+$ and $p\pi^0$ channels, respectively.
For both the experimental and the theoretical values, the integration
  starts at the lowest measured photon energy $E_{\gamma,{\mathrm {min}}}$, that is,  
  $0.175$~GeV fo $n\pi^+$  and  $0.158$~GeV for $p\pi^0$.
Only the statistical errors are shown.}
\label{running}
\end{center}
\end{figure}
In order to check the validity of this estimation,
we evaluated the experimental ``running integral'' $I$
for the FSP $\bar \gamma_0$,
\begin{eqnarray}
\label{exprun}
I(E_{\gamma})= \frac{1}{4\pi^2}\,
\int_{E_{\gamma,\,{\mathrm {min}}}}^{E_{\gamma}}\,\frac{\sigma_{1/2}(\nu)-\sigma_{3/2}(\nu)}{\nu^5}
\,d\nu\,,
\label{eq:running}
\end{eqnarray}
where the lower integration limit $E_{\gamma,\,{\rm {min}}}$
corresponds to the lowest measured photon-energy bin and the upper
integration limit is taken as the running variable.\\
Figure~\ref{running}a) shows the data points based on (i) our evaluation of the helicity dependent
$n\pi^+$ total cross section data in the energy range
$0.175~{\rm {GeV}}\leq E_{\gamma} \leq 0.195~{\rm {GeV}}$~\cite{Ahrens:2004pf} and (ii) the
published helicity-dependent $n\pi^+$ total cross section data in the energy range
$0.195~{\rm {GeV}}\leq E_{\gamma} \leq 0.450~{\rm {GeV}}$~\cite{Ahrens:2004pf}.
The data are compared
with the predictions of four multipole analyses as indicated in the figure. The good agreement between
the experimental data and the results of the HDT and DMT models for ${E}_\gamma < 250$~MeV gives
reasonable confidence in our extrapolation to the lower energies. However, the $1/{\nu}^5$ weight
in the integrand of Eq.~(\ref{eq:running}) clearly enhances the contribution of the unmeasured threshold
region relative to the high energy region. In performing the integrals,
it is also important to consider the isospin breaking effects, at least via
the pion mass splitting.
The pion masses used in our extrapolation to threshold are
$m_{\pi^0}=134.98$~MeV and  $m_{\pi^\pm}=139.57$~MeV for the neutral and charged pions, respectively.
The results for the $\gamma p \to p \pi^0$ channel are shown in Fig.~\ref{running}b) by
comparison of the multipole predictions with the  helicity-dependent data in the range
from  E$_{\gamma}=(0.172 \pm 0.014)$~GeV to E$_{\gamma}=0.45$~GeV~\cite {Ahrens:2004pf}.
Because the measured $p \pi^0$ data point below 0.2~GeV gives only
a very small contribution to the integrals but have a large statistical error, we choose to evaluate
the $p \pi^0$ contribution of this energy range with the HDT analysis.
This choice minimizes the overall (statistical+systematic) error related
to the integral evaluation.\\
\begin{figure}
\begin{center}
\epsfig{file=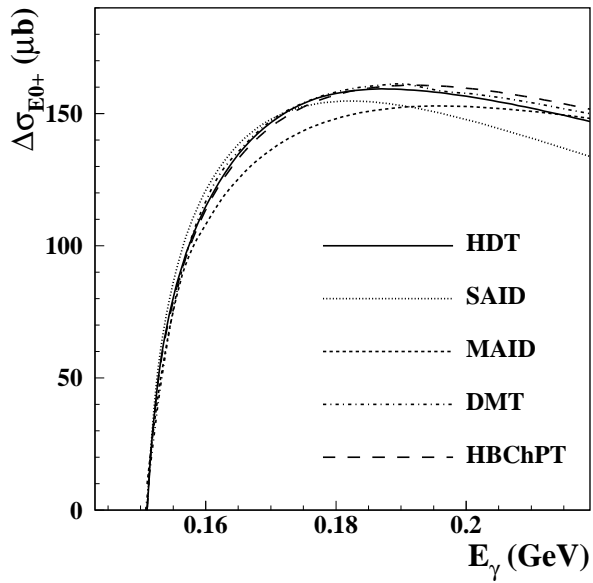,width=0.5\columnwidth}
\caption {The contribution  of the $E_{0+}$ multipole for the charged-pion channel to
$\Delta\sigma=(\sigma_{1/2}-\sigma_{3/2})$ obtained from the HDT~\cite{Hanstein:1997tp},
SAID09~\cite{Arndt:2002xv}, MAID07~\cite{Drechsel:2007if}, and DMT~\cite{DMT:2001} analyses
as well as the predictions of HBChPT~\cite{Fearing:2000}.}
\label{e0p}
\end{center}
\end{figure}
Concerning the unmeasured contribution of
the $\gamma p \rightarrow n\pi^+$ channel below 0.175~GeV, we observe that
the $S$-wave multipole $E_{0+}$ plays an overwhelming role in the threshold region.
An accurate knowledge of this multipole is mandatory
in order to obtain a reliable estimate of this low-energy contribution.
Fortunately, the value of $E_{0+}$ at
threshold is known rather precisely through the low-energy theorem for charged pion photoproduction.
However, current predictions differ with respect to the slope of $E_{0+}$ in the threshold region.
Figure~\ref{e0p} displays the contribution of the $E_{0+}$ multipole for the $n\pi^+$ channel to
the helicity-dependent cross section $\Delta\sigma$ in the threshold region. The results according
to HDT,  SAID09, MAID07, and DMT are compared to the analysis of Ref.~\cite{Fearing:2000}.
The latter work extends the results of low-energy theorems for threshold pion
photoproduction within the framework of HBChPT by fitting the unknown low-energy constants
appearing at ${\cal {O}}(p^3)$ to data on radiative pion capture~\cite{Salomon:1983xn} and
pion photoproduction~\cite{Korkmaz:1999sg}.
\\
Figure~\ref{e0p} shows that the predictions
are in good agreement for $E_{\gamma}\leq 0.175$~GeV except for the MAID results.
The different behavior of MAID is due to the fact that the unitarization involving the $S_{11}(1535)$ resonance
underestimates the size of pion rescattering near threshold. We have therefore evaluated the
contribution of the $n\pi^+$ channel below $E_{\gamma} =175$~MeV using the HDT multipoles.
The associated systematic error has been estimated by comparison with the predictions of the other models excluding
MAID.
\subsection{Discussion of the results}

\begin{figure}[ht]
\epsfig{file=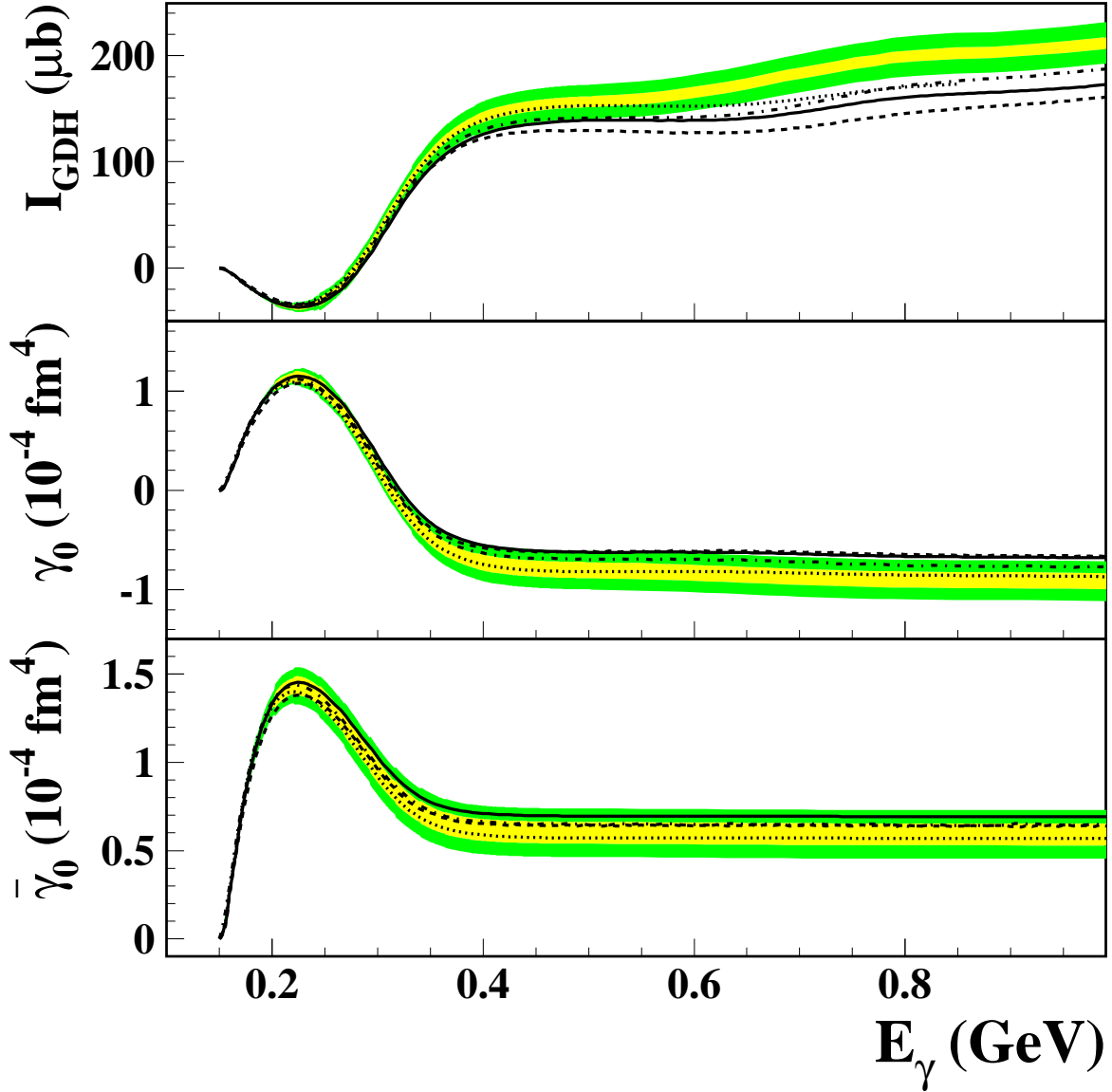,  width=\columnwidth}
\caption{(Color online) The running integrals for the GDH sum rule, $I_{\rm {GDH}}$ (top panel),
the leading FSP $\gamma_0$ (center), and the
next-to-leading FSP $\bar\gamma_0$ (bottom)
as function of the upper limit of integration.
The light grey (yellow) and dark grey (green) bands show the
statistical $(\pm$ sd) and systematic ($\pm \Delta/2$)
uncertainties, respectively, which include both the experimental
errors and the estimated model dependence of the pion photoproduction 
multipoles.
The lines show the predictions based on the pion photoproduction multipoles of
HDT (solid), SAID (dotted), MAID (dashed), and DMT (dash-dotted).}
\label{fig1}
\end{figure}

\begin{figure}[ht]
\begin{center}
\epsfig{file=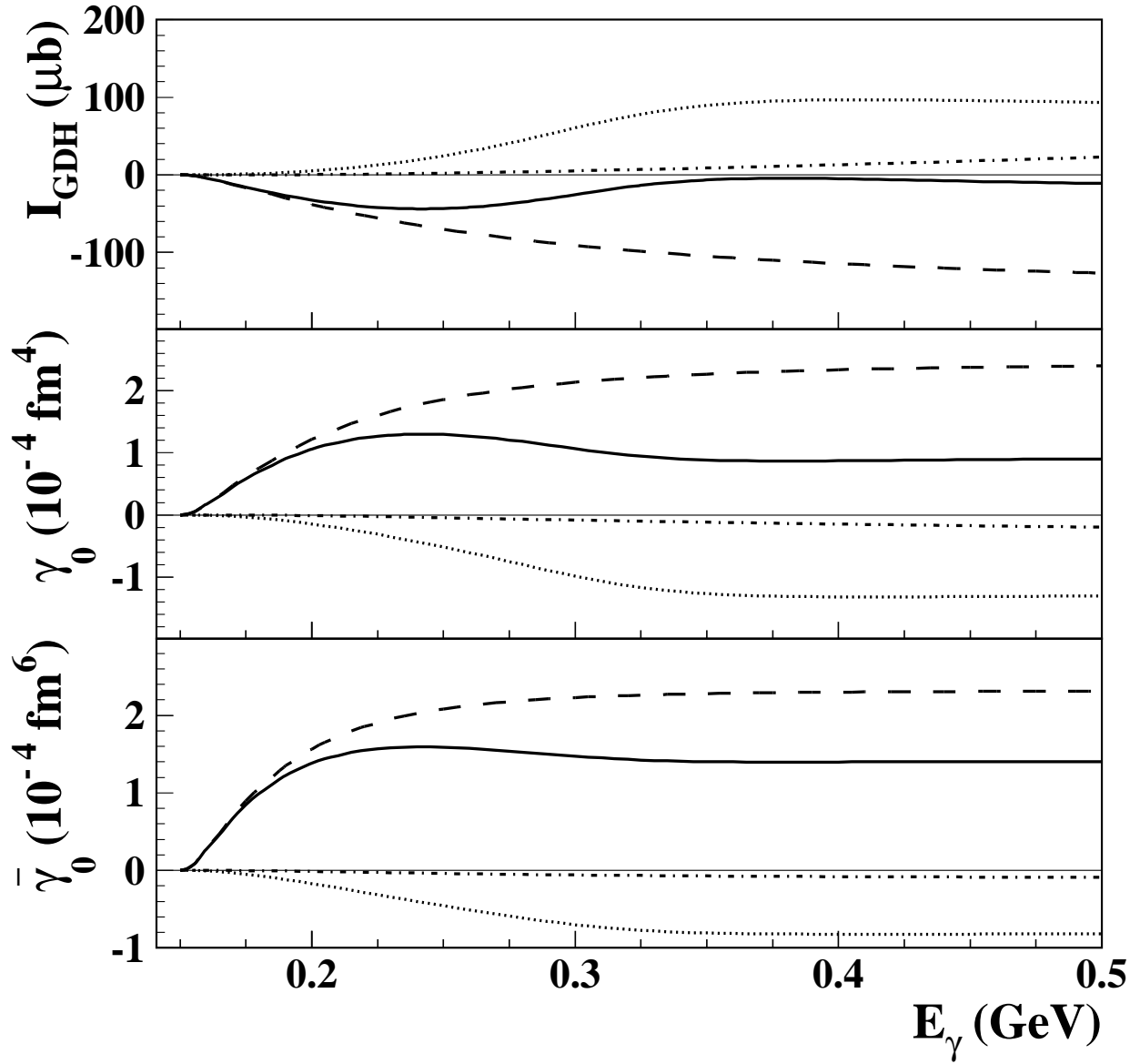,  width=\columnwidth}
\end{center}
\caption{The multipole decomposition of the running integrals for the $n\pi^+$ channel
as obtained from the HDT model. The lines show the contributions
of $S$ waves (dashed), $P$ waves (dotted), and higher partial waves (dashed-dotted) as well as the total result (solid).}
\label{fig3}
\end{figure}

\begin{figure}[ht]
\epsfig{file=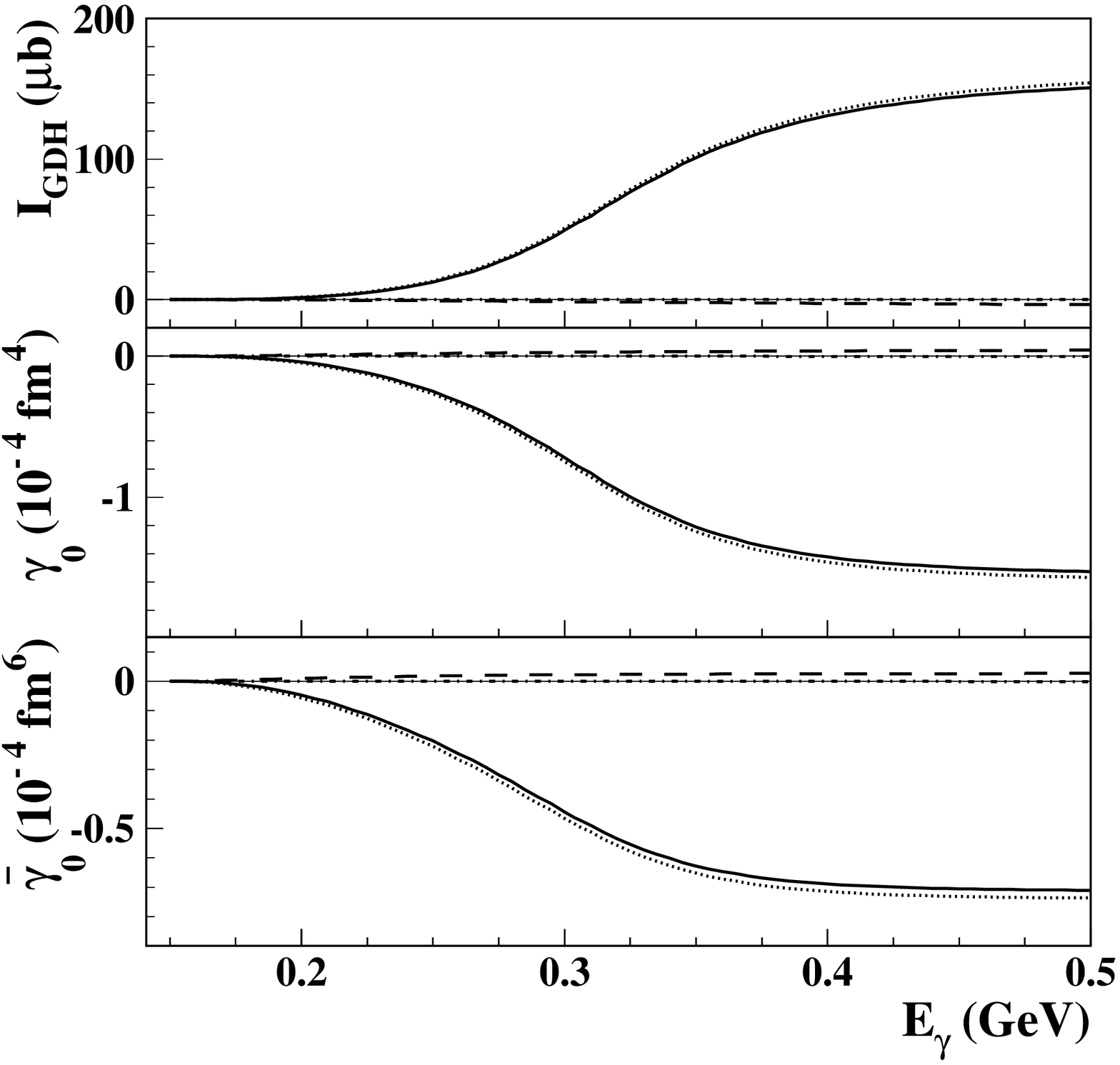,  width=\columnwidth}
\caption{The multipole decomposition of the running integrals for the $p\pi^0$ channel
as obtained from the HDT model. The lines show the contributions
of $S$ waves (dashed) and $P$ waves (dotted) as well as the total result (solid).}
\label{fig2}
\end{figure}

Table~\ref{exper} summarizes the measured and evaluated contributions of the different
energy regions to the GDH sum rule, the leading FSP, and the subleading FSP  as well as
the statistical and systematic errors. Statistical errors are given in
standard deviation (sd) units while systematic ones are given in half interval ($\Delta$/2)
units.
For the GDH sum rule and the leading FSP $\gamma_0$,
the contributions above 2.91~GeV are taken from the estimates given in Ref.~\cite{Drechsel:2008},
whereas such high-energy contributions can be safely neglected for $\bar
\gamma_0$.
\begin{table*}
{\small
\hfill{}
\begin{tabular}{|cc|ccc|ccc|ccc|}
\hline
channel & energy & GDH    & stat. & syst.       & $\gamma_0$ & stat. & syst. & $\bar\gamma_0$ & stat. & syst. \\
        &(GeV)   &($\mu$b)& (sd)  & ($\Delta$/2)&
($10^{-4}$ ${\rm {fm}}^4$) & (sd)&($\Delta$/2)&($10^{-4}$ fm$^{6}$)& (sd)  &($\Delta$/2)\\
\hline
$n\pi^+$ & $\leq 0.175$ &
 $-16$   &              &$\pm 1$    &
$+0.56$   &              & $\pm 0.01$&
$+0.83$    &              & $\pm 0.02$ \\
$p\pi^0$ & $\leq 0.195$    &
1        &    & $\pm0.1$ &
$-0.03$  &    &$\pm 0.05$&
$-0.04$  &    & $\pm 0.01$ \\
$n\pi^+$ & $0.175 - 0.195$ &
$-14$    &$\pm 1$     & $\pm 1$   &
$+0.40$   &$\pm 0.02$  & $\pm 0.02$&
$+0.46$    & $\pm 0.02$ & $\pm 0.02$ \\
$\gamma p\rightarrow X$ & $0.195 - 2.91$ &
253      &$\pm 6$     & $\pm 13$&
$-1.85$  & $\pm 0.07$ & $\pm 0.05$&
$-0.65$  & $\pm 0.07$ & $\pm 0.05$ \\
$\gamma p \rightarrow X$ & $ > 2.91 $ &
 $-14$   &            &$\pm 2$  &
0.01     &            &         &
& &\\
\hline
total  &           &
210    &$\pm 6$    &$\pm 14$&
$-0.90$&$\pm 0.08$ &$\pm 0.11$&
$+0.60$ & $\pm 0.07$ & $\pm 0.07$ \\
\hline
\end{tabular}}
\hfill{}
\caption{Contributions from different photon energy regions to the GDH integral $I_{\rm {GDH}}$,
the leading FSP $\gamma_0$, and the subleading FSP $\bar \gamma_0$ defined by Eqs.~(\ref{eq:GDH})
and (\ref{eq:FSP}) as well as the associated statistical and systematic errors.
The GDH sum rule predicts 204 $\mu b$. See text for further
information.}
\label{exper}
\end{table*}
\newline
\noindent
\begin{table*}
{\small
\hfill{}
\begin{tabular}{|c||c|c||c|c|c|c||c|}
\hline
& $\gamma_0$ & $\bar\gamma_0$ & $\tilde\gamma_0$  & recoil \\
& ($10^{-4} \, \rm{fm}^4$) & ($10^{-4} \, \rm{fm}^6$)& ($10^{-4} \, \rm{fm}^6$)& ($10^{-4} \, \rm{fm}^6$) \\
\hline
\hline
${\cal O}(p^3)$~\cite{Holstein:1999uu}        &  4.6     & 4.23 & 5.15 & 0.92 \\
\hline
${\cal O}(\epsilon^3)$~\cite{Holstein:1999uu} &  2.9     & 3.65 & 4.23 & 0.58 \\
\hline
dispersion~\cite{Hanstein:1997tp}             &  $-0.69$ & 0.69 & 0.56 & $-0.14$ \\
\hline
our evaluation &  $-0.90 \pm 0.08$  $\pm 0.11$& $0.60\pm 0.07\pm 0.07$
& $0.42 \pm0.09 \pm 0.09$ & $0.18 \pm 0.02 \pm 0.02$ \\
\hline
\end{tabular}}
\hfill{}
\caption{The FSPs of the proton. The leading FSP $\gamma_0$
appears as coefficient of the amplitude at third order in both the lab energy $\nu$ and the c.m. energy $\omega$.
The next-to-leading FSP, $\bar \gamma_0$ in the lab system and $\tilde \gamma_0$ in the c.m. system, appear at
fifth order in the respective energies. They differ by the recoil term given in the last column of the table.
The predictions are obtained from HBChPT at ${\cal O}(p^3)$ and ${\cal O}(\epsilon^3)$~\cite{Holstein:1999uu} as well
as dispersion theory based on the HDT multipoles of the one-pion channels.}
\label{tab-II}
\end{table*}
In Table~\ref{tab-II} we compare our results for the FSPs $\gamma_0$ and
$\bar\gamma_0$ to (i) predictions obtained in HBChPT at ${\cal O}(p^3)$ and
at ${\cal O}(\epsilon^3)$ in the small scale expansion (SSE) including the $\Delta(1232)$ as an
explicit degree of freedom~\cite{Holstein:1999uu} as well as
(ii) results based on dispersion relations using the pion photoproduction multipoles
of HDT up to 500 MeV and extended by the SAID multipoles to the higher energies. The FSPs found from
our analysis are in good agreement with the results obtained by dispersion relations.
However, the FSPs are considerably overestimated by lowest order HBChPT.
Further works on the leading spin polarizability $\gamma_0$ resulted in the following
values (all in units of 10$^{-4}$ fm$^4$):
$-3.9$ to ${\cal O}(p^4)$ in HBChPT~\cite{VijayaKumar:2000pv},
2.0 to  ${\cal O}(\epsilon^3)$ in SSE~\cite{Hemmert:1997tj},
and $-0.9$ at NNLO in manifestly Lorentz-covariant ChPT~\cite{Lensky:2009uv}.
The dependence of the dispersive predictions on the one-pion photoproduction multipoles is
illustrated in Table~\ref{tab-III}. The solution labeled HDT corresponds to the HDT multipoles
as given over the energy range $\nu_0 <E_{\gamma} < 0.5$~GeV,
supplemented by the SAID parametrization for $0.5~{\rm {GeV}} <E_\gamma < 3$~GeV.
The MAID07 and SAID09 predictions are obtained from the most recent parameterizations of
these models for $\nu_0 <E_{\gamma} < 3$~GeV. Finally, the solution labeled DMT is based
on the DMT multipoles in the range $\nu_0 <E_{\gamma} < 1.7$~GeV, supplemented by the
SAID09 parametrization in the range $1.7~{\rm {GeV}}<E_{\gamma} < 3 $~GeV.\\
Beyond the one-pion channel, we estimated the contribution from multi-pion intermediate states
by inelastic decay of $\pi N$ resonances as detailed in Ref.~\cite{Drechsel:1999rf}. The inelastic
contributions are assumed to have the same helicity structure as the one-pion channels.
In this approximation, we first calculate the resonant part of the pion photoproduction
multipoles using the Breit-Wigner parametrization of Ref.~\cite{Amsler:2008zb} and then scale
this contribution by a suitable factor to include the inelastic decay of the resonances.
As shown in Table~\ref{tab-III}, this procedure yields a multi-pion contribution of
$(20 \pm 2)\, \mu$b to the GDH sum rule. On the other hand, we have evaluated this contribution
on the basis of recently published data for the helicity dependence of the $\gamma p \to N\pi\pi$
channels in the energy range $325$~MeV $< E_\gamma< 800$~MeV~\cite{pi0pip:2003,pi0pi0:2005,pimpip:2007}.
>From the data in the limited energy range covered by the experiments, we have obtained
a two-pion contribution of $(39 \pm 1 \pm 3)\, \mu$b, about twice the value of the model predictions.
The difference between the data and the predictions has its origin in the strong two-pion
contribution observed in the energy range $500<E_\gamma<700$ MeV, which can not be described by the resonance model.
In view of this discrepancy, an improved theoretical description of the multi-pion
continuum states is prerequisite to describe the GDH sum rule in a quantitative way.
As may be expected from the energy-weight factors in the respective integrals,
Table~\ref{tab-III} shows only small multi-pion effects for the FSPs.
\newline
\noindent
Figures~\ref{fig1}-\ref{fig2} display the running integrals
for the GDH integral $I_{\rm {GDH}}$, the leading FSP $\gamma_0$ and the subleading FSP $\bar\gamma_0$. These integrals
are defined as in Eq.~(\ref{eq:running}), with the appropriate weight factors in the denominator.
Figure~\ref{fig1} compares our result with the predictions of several multipole analyses.
The shaded bands represent the statistical and systematic errors associated with our analysis, which includes
both the experimental errors and the model errors of the extrapolation into the unmeasured regions.
The model predictions lie within the error bars of our analysis, except for the running GDH integral above 0.5~GeV.
In the latter case the model predictions stay below the data by about $30\,\mu$b, mostly because of
the missing multi-pion strength.
\begin{table}
\begin{tabular}{|c||c|c|c|}
\hline
     & GDH       & $\gamma_0     $   & $\bar \gamma_0$  \\
     & ($\mu$b)  & ($10^{-4}$ fm$^4$) & ($10^{-4}$ fm$^6$)\\
\hline
HDT  & 177 (196) & $-0.67 (-0.69)$   & 0.69 (0.69) \\
MAID & 163 (182) & $-0.65 (-0.67)$   & 0.65 (0.64) \\
SAID & 191 (208) & $-0.86 (-0.88)$   & 0.57 (0.57) \\
DMT  & 185 (205) & $-0.76 (-0.78)$   & 0.64 (0.64) \\
\hline
\end{tabular}
\caption{
Results for the GDH sum rule,
the leading FSP $\gamma_0$, and the subleading FSP $\bar\gamma_0$ as obtained
from the pion photoproduction multipoles of HDT, MAID, SAID, and DMT. The values in
bracket include the multi-pion contributions.}
\label{tab-III}
\end{table}

The multipole decompositions of the $n \pi^+$ and $p \pi^0$ channels are shown in
Figs.~\ref{fig3} and \ref{fig2}, respectively. The $n\pi^+$  channel is
characterized by a strong competition between the $E_{0^+}$ multipole above threshold and the $M_{1^+}$ near the
$\Delta (1232)$ resonance,
whereas the neutral pion channel is almost completely described by $\Delta (1232)$ resonance effects.\\
\begin{figure}[ht]
\begin{center}
\epsfig{file=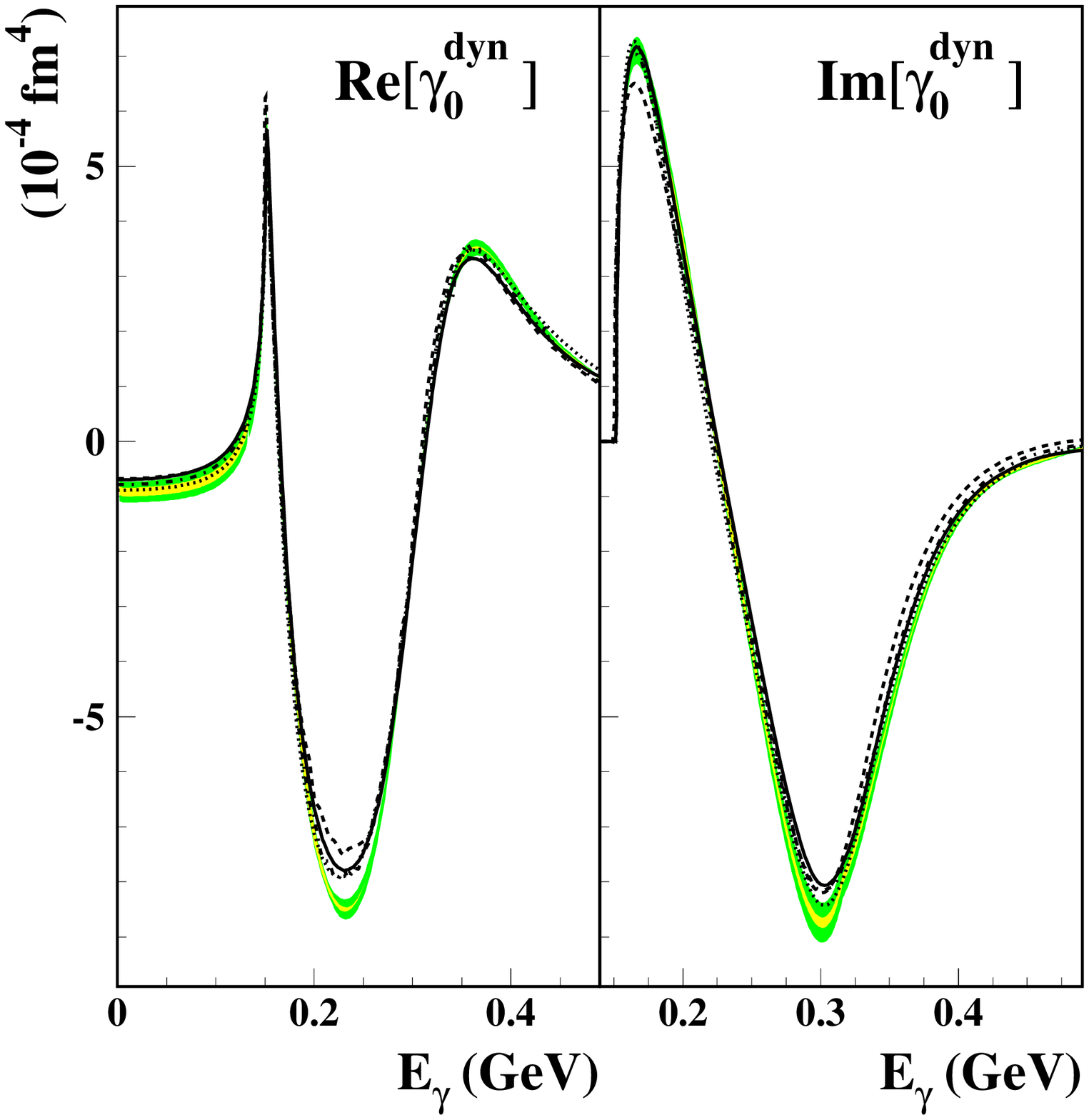,  width=\columnwidth}
\end{center}
\caption{(Color online)
The real (left panel) and imaginary (right panel) parts of the dynamic FSP
$\gamma_0^{\rm {dyn}}$ as function of the photon lab energy $E_{\gamma}$.
The light grey (yellow) and dark grey (green) bands show the
statistical $(\pm$ sd) and systematic ($\pm \Delta/2$)
uncertainties, respectively, which include both the experimental
errors and the estimated model dependence of the pion photoproduction multipoles.
The lines show the predictions based on the pion photoproduction multipoles of
HDT (solid), SAID (dotted), MAID (dashed), and DMT (dash-dotted).}
\label{fig4}
\end{figure}
The real and imaginary parts of the dynamic FSP $\gamma_0^{\rm {dyn}}$, as defined by Eq.~(\ref{eq:dyn-FSP}),
are displayed in Fig.~\ref{fig4}. 
Up to the lowest threshold at $E_{\gamma}=\nu_0$, the real part
can be expanded in a power series according to the LEX, Eq.~(\ref{eq:dyn-FSP-LEX}). Of course, the convergence
of the series deteriorates with increasing energy. Up to  
$E_{\gamma}=50$~MeV, the forward scattering amplitude
is described by the leading polarizability $\gamma_0$ within an accuracy 
of 10\%. In order to obtain the
same precision at 80~MeV, also $\bar \gamma_0$ must be included, and 
already four terms of the series are necessary
at 100~MeV.  As shown in Fig.~\ref{fig4}, the onset of $S$-wave pion production
at $E_{\gamma}=\nu_0$ leads to a strong cusp effect in the real part. The rapid increase of the dynamic FSP
from its static value $-0.90\cdot 10^{-4}$ ${\rm {fm}}^4$ at $E_{\gamma}=0$ to about $6\cdot 10^{-4}$ ${\rm {fm}}^4$
at pion threshold clearly shows the necessity to analyze Compton scattering within the framework of
dispersion analysis. In particular, such an approach is prerequisite in order to determine the  spin
polarizabilities, which yield significant contributions to the cross section only for photon energies above 80~MeV.
Near $E_{\gamma}\approx 300$~MeV, the dynamic FSP is dominated by the resonance structure of the $\Delta (1232)$.
Except for the minimum of ${\rm {Re}}[\gamma_0^{\rm{dyn}}]$ near $E_{\gamma}=0.23$~GeV, our analysis is in
very good agreement with the predictions of the shown models. The deviation in the minimum is not too surprising,
because this comes about by a delicate balance between the $S$-wave background
and the low-energy tail of the $\Delta (1232)$ resonance.
\section{Conclusions}
The polarizabilities of the nucleon are elementary properties providing
stringent tests for theoretical approaches to hadron physics, such as chiral
perturbation and lattice gauge theories. The (scalar) electric and
magnetic dipole polarizabilities have been determined quite precisely by comparing
Compton scattering data to the model-independent low-energy expansion at
sufficiently low photon energies. However, this procedure fails for the ``missing''
four spin polarizabilities because of their suppression in the cross sections at low photon energies.
In order to extract these polarizabilities from the data, it is prerequisite to
(i) measure polarized Compton scattering in the energy range of about 150-300~MeV and
(ii) analyze the data within a theoretical framework that describes the physics in this range
sufficiently well. Such a theoretical basis is provided by dispersion relations with
data on pion photoproduction as input. The critical question is, of course, the model-dependence
of such an analysis, in particular with regard to multi-pion and heavier meson production
as well as unknown high-energy tails.
In order to study these questions, we have constructed the spin-dependent part of the forward Compton amplitude
using the recently measured helicity-dependent photoabsorption cross sections.
Although this ``experimental'' forward spin amplitude needs some theoretical input for
the unmeasured region near pion production threshold, the associated systematic
errors are small because of the (model-independent) low-energy theorem for charged pion photoproduction
at threshold.
We have compared our findings to predictions based on pion photoproduction multipoles
given by chiral perturbation theory as well as phenomenological models.
This comparison includes the Born term given by the GDH integral and the dispersive contributions
related to the nucleon's excitation spectrum, as expressed by the leading and higher order FSPs
as well as the full (dispersive) forward spin amplitude or ``dynamic'' FSP. As demonstrated by the
``running'' integrals for these observables, the theoretical predictions and the results based
on the experimental data agree quite well for photon energies below 300~MeV. At the higher energies we find
deviations up to 10-20\%, mostly due to the modeling of the helicity structure of multi-pion production.
Because of the suppression by energy-dependent weight factors, the high-energy contribution to
the next-to-leading and higher order spin polarizabilities is much reduced. It is therefore a viable strategy to
analyze future polarization experiments by (i) 
treating the leading polarizabilities as free parameters and (ii) fixing the
higher order polarizabilities by subtracted dispersion integrals based on the pion photoproduction multipoles.
We are confident that such experiments will advance our understanding of a basic property of the
nucleon: the response of its spin structure to an applied electromagnetic field.

\section*{Acknowledgements}
The authors are grateful to  J\"{u}rgen Ahrens, Hans-J\"{u}rgen Arends  and 
Lothar Tiator for advice and discussions.
This work was supported by the Deutsche Forschungsgemeinschaft (SFB 443)
and the Research Infrastructure Integrating Activity
``Study of Strongly Interacting Matter'' (HadronPhysics2,
Grant Agreement n. 227431) under the 7th framework programme of the
European Community.
\end{document}